\documentclass[journal=jacsat,manuscript=article]{achemso}

\usepackage[version=3]{mhchem} 
\usepackage{color,soul}
\usepackage{charter} 

\author{Peichen Zhong}
\affiliation[UCB]
{Department of Materials Science and Engineering, University of California Berkeley, Berkeley CA 94720, USA}
\alsoaffiliation[LBNL]
{Materials Sciences Division, Lawrence
Berkeley National Laboratory, Berkeley CA 94720, USA}
\author{Sunny Gupta}
\affiliation[UCB]
{Department of Materials Science and Engineering, University of California Berkeley, Berkeley CA 94720, USA}
\alsoaffiliation[LBNL]
{Materials Sciences Division, Lawrence
Berkeley National Laboratory, Berkeley CA 94720, USA}
\author{Bowen Deng}
\affiliation[UCB]
{Department of Materials Science and Engineering, University of California Berkeley, Berkeley CA 94720, USA}
\alsoaffiliation[LBNL]
{Materials Sciences Division, Lawrence
Berkeley National Laboratory, Berkeley CA 94720, USA}
\author{KyuJung Jun}
\affiliation[UCB]
{Department of Materials Science and Engineering, University of California Berkeley, Berkeley CA 94720, USA}
\alsoaffiliation[LBNL]
{Materials Sciences Division, Lawrence
Berkeley National Laboratory, Berkeley CA 94720, USA}
\author{Gerbrand Ceder*}
\email{gceder@berkeley.edu}
\affiliation[UCB]
{Department of Materials Science and Engineering, University of California Berkeley, Berkeley CA 94720, USA}
\alsoaffiliation[LBNL]
{Materials Sciences Division, Lawrence
Berkeley National Laboratory, Berkeley CA 94720, USA}

\title{Effect of cation-disorder on lithium transport in halide superionic conductors}
\abbreviations{}

\begin{document}


\clearpage
\begin{abstract}
Li$_2$ZrCl$_6$ (LZC) is a promising solid-state electrolyte due to its affordability, moisture stability, and high ionic conductivity. We computationally investigate the role of cation disorder in LZC and its effect on Li-ion transport by integrating thermodynamic and kinetic modeling. The results demonstrate that fast Li-ion conductivity requires Li/vacancy disorder, which is dependent on the degree of Zr disorder. The high temperature required to form equilibrium Zr-disorder precludes any equilibrium synthesis processes for achieving fast Li-ion conductivity, rationalizing why only non-equilibrium synthesis methods, such as ball milling, lead to good conductivity. Our simulations show that Zr disorder lowers the Li/vacancy order-disorder transition temperature, which is necessary for creating high Li diffusivity at room temperature. These insights raise a challenge for the large-scale production of these materials and the potential for the long-term stability of their properties.
\end{abstract}

\begin{figure*}[t]
\centering
\includegraphics[width = \linewidth]{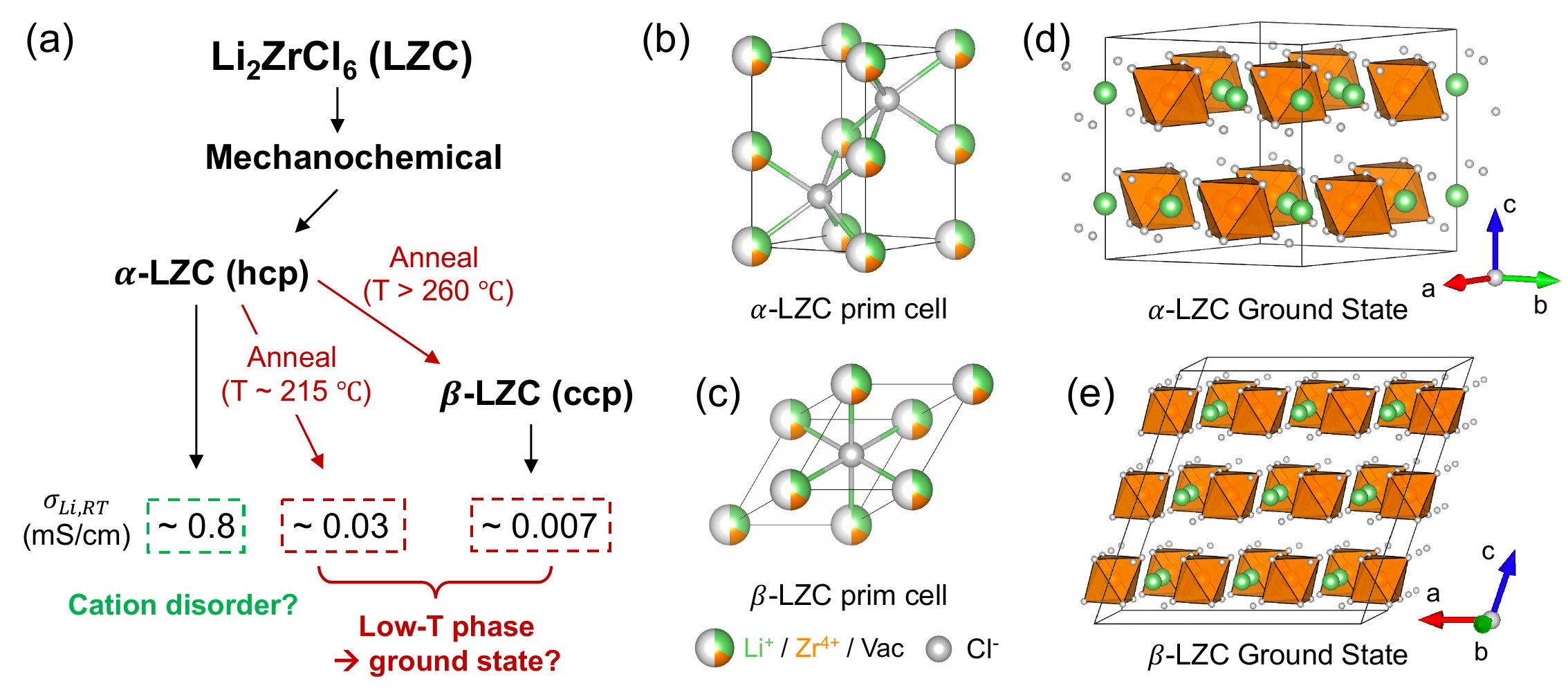}
\caption{(a) A summary of the synthesis paths and ionic conductivities of LZCs based on the experimental results reported by \citet{Wang2021_LZC} (b)-(c) The primitive cells used for cluster expansion models of $\alpha$-LZC (hcp) and $\beta$-LZC (fcc). The green/orange/white spheres represent the cation site that can be occupied by Li/Zr/vacancies, respectively. The grey sphere represents the anion site with Cl occupancy. The ground-state structures of (d) $\alpha$-LZC and (e) $\beta$-LZC.}
\label{fig:GS_prim}
\end{figure*}

High-Li-ion-conductivity solid electrolytes are essential for realizing all-solid-state batteries (ASSBs), which offer the potential for increased volumetric energy density and improved safety compared to conventional liquid-electrolyte-based batteries \cite{Janek2016_future, xia2019practical, sun2020promising}. Chloride-based Li-ion solid electrolytes, with the general formula Li$_x$M$_y$Cl$_6$, where M represents a transition metal and in which Cl anions form a close-packed framework, have become of interest as catholytes in ASSBs \cite{li2020progress, wang2022prospects, kwak2022emerging, Helm2021_InZr, Schlem2021_LYX, Zhou2022_lindar, Zhou2023_HoLu}. Many of these electrolytes exhibit high room-temperature (RT) Li-ion conductivity ($>1$ mS/cm), good interfacial stability with cathode materials, and high oxidation stability \cite{asano2018solid, wang2019lithium, xiao2020understanding, wang2022prospects, Zhou2022_lindar}, consistent with early theory predictions \cite{Richards2016}. Significant challenges still exist before these materials can be commercialized in practical cells. Some of the halide conductors contain low-abundance metals (e.g., Sc/Y/Tb/Lu). In addition, many of the interesting halides are unstable in ambient conditions and their ionic conductivity significantly degrades by exposure to moisture \cite{Zhu2020_air, li2020progress}. Recently, the Zr-based compounds (e.g., Li$_2$ZrCl$_6$, LZC) and their doped variants have demonstrated the potential to overcome these challenges \cite{Wang2021_LZC, kwak2021new, li2022_nanoletter, kwak2022li+_CEJ, Helm2021_InZr, li2023high_CEJ, chen2023unraveling, Shi2024_ACS_sustain}. 
LZC exhibits distinct polymorphism with its structure depending on the synthesis method as summarized in Figure \ref{fig:GS_prim}a. The mechanochemical synthesis produces $\alpha_h$-LZC (hexagonal close-packed anions, hcp), whereas the $\beta_c$-LZC (cubic close-packed anions, ccp) can form by annealing $\alpha_h$-LZC at temperatures above 260 $^\circ$C.
While the ball-milled $\alpha_h$-LZC exhibits RT Li-ion conductivity of approximately 0.8 mS/cm \cite{Wang2021_LZC}, the annealed LZC in either the $\alpha$- or $\beta$-phase display significantly lower RT Li-ion conductivity \cite{Wang2021_LZC}. This has led to the hypothesis that some disorder is critical to achieve high ionic conductivity. Mechanochemical ball-milling as a non-equilibrium synthesis method is capable of inducing high-energy cation-disordered configurations \cite{Kitchaev2018_EES, Lun2019_Chem}. 
\citet{Schlem2020_disorder} demonstrated that in  Li$_3$YCl$_6$ and  Li$_3$ErCl$_6$ mechanochemical synthesis induces cation defects by creating disorder on the sites occupied by Y/Er and that this results in a reduced activation energy and enhanced ionic conductivity.
Using molecular dynamics simulations, \citet{Wang2023_frustration} revealed that the Li sublattice disorders above $T_c=425$ K in Li$_3$YCl$_6$, and that the associated broadening of the Li site energies creates fast Li-ion diffusion. Other approaches such as introducing stacking faults \cite{Sebti2022_stacking} and tunning Li/metal stoichiometry to create percolation pathways \cite{Liang2020_disorder, Yu2023_science} have also been proven to enhance Li-ion conductivity.

In this study, we seek to explain the correlation between synthesis route, cation disorder, and ion transport in LZC systems through \textit{ab initio} modeling. We demonstrate that, in equilibrium, Li- and Zr-disorder occur at very different temperatures. Our results suggest that facile Li disorder only emerges at RT in the presence of Zr disorder, but that equilibrium Zr disorder only occurs at very high temperatures. In the low-temperature equilibrium states in which Zr is well-ordered, Li/vacancy intersite exchanges are thermodynamically limited, leading to very low conductivity. The finding that at low temperature Li disorder is driven by Zr disorder is further confirmed by molecular dynamics simulations, in which we find a significant improvement in Li ionic diffusivity with the presence of Zr disorder. Our investigation demonstrates that Zr disorder is essential to achieve high Li conductivity, but cannot likely be achieved through thermal equilibrium synthesis routes, thereby explaining why extensive ball milling is required for these materials. We believe this to be a challenge for the large-scale production of these materials as well as a potential problem for the long-term stability of their properties.

\begin{figure*}[t]
\centering
\includegraphics[width = \linewidth]{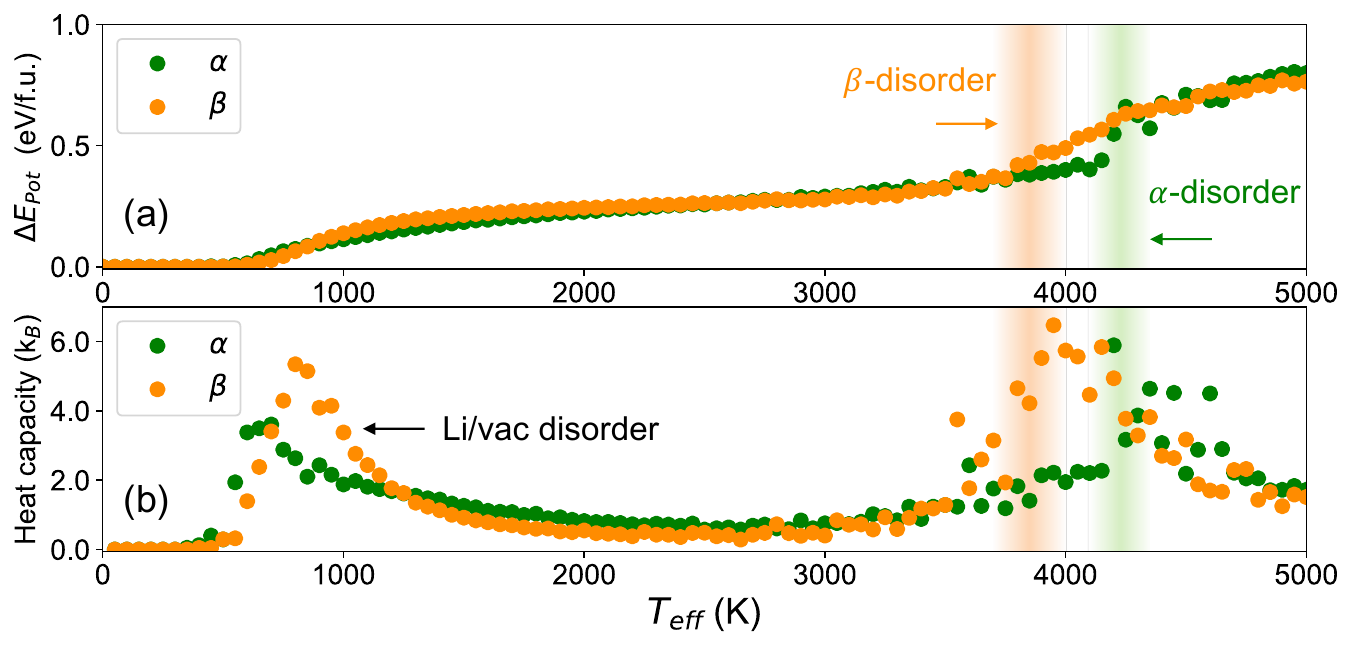}
\caption{(a) The average potential energy above the ground state (\(\Delta E = E(T_{\text{eff}}) - E_{\text{GS}}\)) sampled full equilibrium CE-MC simulations.  The green/orange dots represent $\alpha$/$\beta$-LZC, respectively. The shaded regions label the increase of potential energy as an indication of the Zr disorder. (b) The calculated heat capacity $C_v$. The first rise in $C_v$ corresponds to the transition to Li/vacancy disorder, and the second rise (indicated by the shaded region) corresponds to the transition where Zr disorders.
}
\label{fig:T_MC}
\end{figure*}

To investigate the thermodynamics of cation disorder in LZCs, we conducted cluster expansion (CE) Monte Carlo (MC) simulations to sample the average and fluctuation of the configurational energy at finite temperatures. The CE is a mathematically rigorous approach to expanding the energy of a system in terms of configurational variables (site occupancies) which has been used to study the configurational thermodynamics of materials in which sites can be occupied by multiple cations \cite{Sanchez1984, Barroso-Luque2022_CE_theory, Barroso-Luque2022_smol}, and has been applied to study many battery materials with cation disorder, including cathodes \cite{Wolverton1998_LiCoO2, Zhong2023_PRX} and solid-state electrolytes \cite{Deng2020, Kam2023}, as well as for disorder in metallic systems \cite{VandeWalle2009}. The CE model for $\alpha$/$\beta$-LZC was constructed based on the hcp/fcc primitive cell as shown in Figure \ref{fig:GS_prim}b/c. The expansion coefficients in the CE were fitted to the energies of various cation configurations calculated using density functional theory (DFT), with the standard methodological details provided in the Supporting Information. The ground states (GS) of both $\alpha$- and $\beta$-LZC were found to exhibit a layered structure, as shown in Figure \ref{fig:GS_prim}d--e. Monte Carlo (MC) sampling of the CE energies can be used to thermally equilibrate a system at finite temperatures.  We started from the ground state structures and heated them from $0$ to $5000$ K.  Figure \ref{fig:T_MC}a presents the average potential energy above the ground state \(\Delta E_{\text{pot}} = E(T_{\text{eff}}) - E_{\text{GS}}\), where $T_{\text{eff}}$ is the effective temperature used in MC simulations at which equilibrium atomic configurations were sampled. Figure \ref{fig:T_MC}b shows the heat capacity \(C_v = \sigma^2(E)/ k_BT^2\), where the green/orange dots represent \(\alpha/\beta\)-LZC, respectively. The two significant increases in \(\Delta E_{\text{pot}}\) accompanied by peak signals in \(C_v\) for both \(\alpha/\beta\)-LZC indicate order--disorder phase transitions. The heat capacity peak near 500 K is associated exclusively with Li/vacancy disorder, and no Zr disorder is detected in the MC structures in this temperature range. The subsequent increase in \(\Delta E_{\text{pot}}\) and corresponding peak in \(C_v\) near 4000 K corresponds to the Zr order--disorder phase transition, which is marked by the green/orange shaded region in Figure \ref{fig:T_MC}. Such a high temperature suggests that there is no spontaneous Zr disorder from thermal fluctuations at the typical temperatures accessible during solid-state synthesis. Given these results, it is clear that only high-energy, non-equilibrium synthesis approaches such as mechanochemical ball-milling can achieve the disordered state.

\begin{figure*}[ht]
\centering
\includegraphics[width = \linewidth]{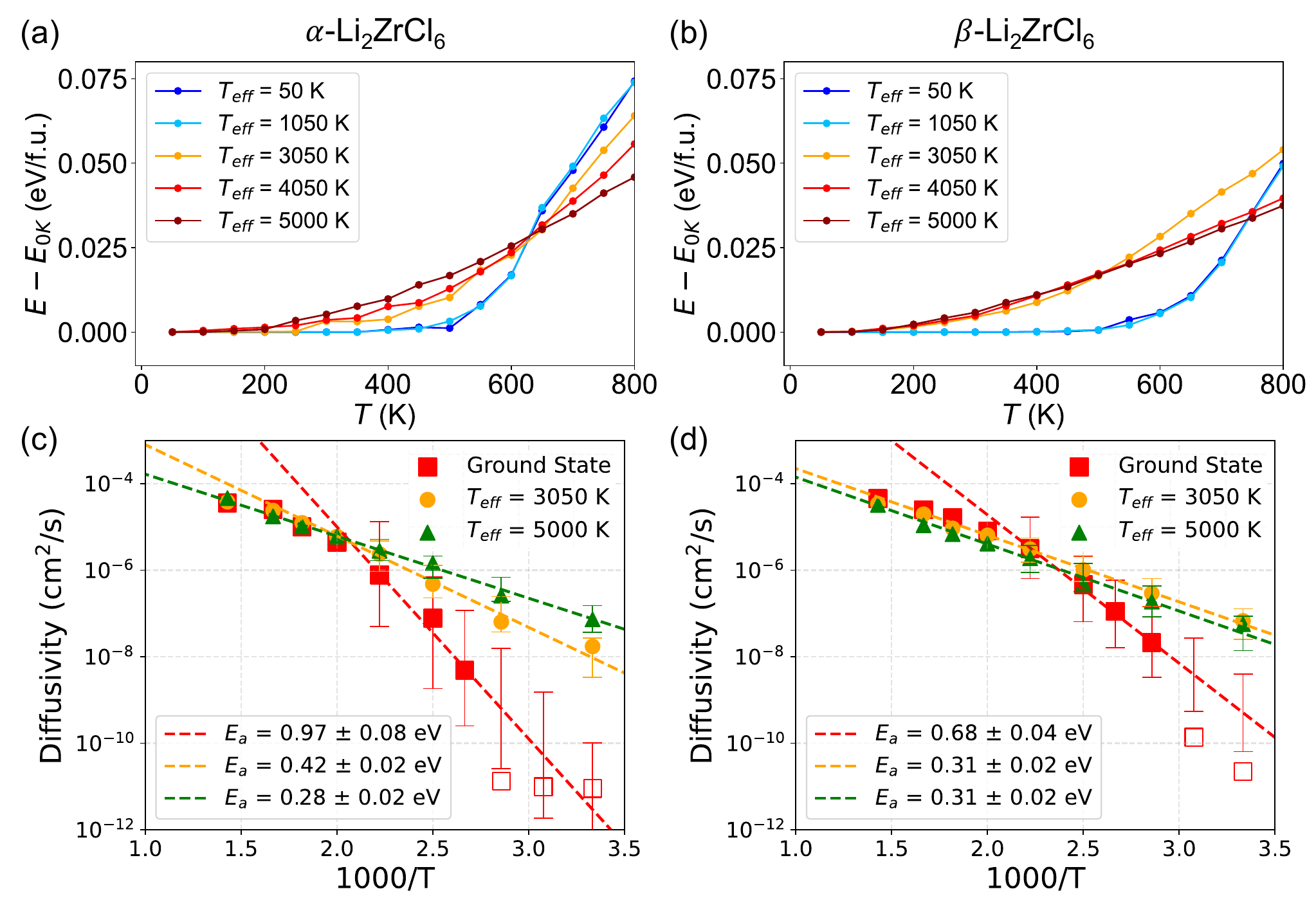}
\caption{(a)-(b) The Li/vacancy configurational energy landscape $\{E_T\}$ sampled by MC annealing with fixed Zr ordering. The color lines represent the average energy sampled by the Li/vacancy degrees of freedom with various Zr orderings obtained by equilibration at  $T_{\text{eff}}$. The rise in $\{E_T\}$ indicates the activation of Li/vacancy inter-site exchanges (disorder). As the $T_{\text{eff}}$ at which the Zr was disordered increases, Li can more easily sample higher energy states. (c)-(d) Arrhenius plots of the MD simulated diffusivities \textit{vs.} temperature from 300 K to 700 K. The MD simulations started from the Zr configurations obtained with MC at $T_{\text{eff}}=0$ K (green triangles), 3050 K (orange dots), and 5000 K (red squares).
The dashed lines represent the activation energy fitted in the low-temperature range ($T<500$ K). A significant non-Arrhenius behavior is observed in Li diffusion in the ground-state structure. The presence of Zr disorder reduces the degree of non-Arrhenius behavior, and the dashed line fitted in low-temperature regions extends well to the high-temperature region.}
\label{fig:landscape_MD}
\end{figure*}

To evaluate how the Li/vacancy site distribution is affected by the presence of Zr disorder, we investigated the energy sampled by the Li/vacancy configurations in a MC simulation when Zr ordering is fixed to that obtained at elevated temperatures. The Zr configuration sampled from MC at $T_{\text{eff}}=3050$ K shows intersite exchange between Li/Zr for both $\alpha$- and $\beta$-LZC, while the Zr configuration above the order--disorder transition (e.g., $T_{\text{eff}}=5000$ K) corresponds to complete cation disorder on the cation sublattices in the hcp and fcc framework (Figure S2). Figure \ref{fig:landscape_MD}a--b shows the average configurational energy, $\{ E_T\}$, sampled by the Li/vacancy degrees of freedom conditioned on various fixed Zr states of disorder (represented by $T_{\text{eff}}$, which is the temperature at which the Zr configurations were obtained).   The $x$-axis represents the temperature at which the Li/vacancy energy was sampled in MC, and the $y$-axis shows the average energy difference above the 0 K configuration. The Li/vacancy energetics for both \(\alpha\)- and \(\beta\)-LZC with  $T_{\text{eff}}=50$ K and 1050 K are almost the same, indicating no Zr disorder occurs in this temperature range. When Zr is well ordered ($T_{\text{eff}}=50$ K or $T_{\text{eff}}=1050$ K), the Li/vacancy disorder initiates at approximately 500 K indicated by the characteristic potential-energy increase. 

For the structures with a high degree of Zr disorder (e.g., \(T_{\text{eff}}=5000\) K), Li already samples higher-energy configurations at low temperatures (200 to 300 K) and no distinct order--disorder transition is observed, consistent with findings in other materials with coupled disorder \cite{Tepesch1995}. This observation indicates that Li/vacancy configurational disorder at RT is made possible by Zr disorder. An analysis of Li intersite exchange energy ($\Delta E_{\text{site}}$) is provided in Figure S3, which finds that the presence of Zr disorder broadens the site energy distribution and increases the density of low $\Delta E_{\text{site}}$.

\clearpage

To obtain quantitative measures of Li diffusivity as a function of Zr disorder, we employed a machine learning interatomic potential (MLIP) to perform molecular dynamics (MD) simulations and investigate Li transport kinetics. By simulating the atomic trajectories, MD is typically used to obtain conductivities in fast-ion conductors through the Green--Kubo formalism. The presence of cation disorder and slow transport kinetics near RT necessitates large-scale, long-time simulations. MLIP is computationally efficient while maintaining close to \textit{ab-initio} accuracy, making it well-suited for this purpose. The MLIP was fine-tuned from the pretrained CHGNet using energies, interatomic forces, and stresses obtained from DFT calculations of various atomic configurations (see Supporting Information for model details) \cite{deng2023_chgnet, Vandermause2022_active}. The Li diffusivity is estimated from the mean-squared displacement over simulation time via $D = \sum_i^N [\Delta \Vec{r}(t)]^2/(6Nt)$ following the empirical error estimation scheme proposed by \citet{He_Zhu_Epstein_Mo_2018}, where $\Vec{r}(t)$ represents the displacement of the $i$-th diffusive atom and $t$ denotes the simulation time. The activation energy is computed by fitting the logarithm of diffusivity \textit{vs.} the inverse of temperature. In the MD simulations, structures with partial Zr disorder were obtained from MC simulations at $T_{\text{eff}}=0/3050/5000$ K, and Zr ions do not migrate so Li diffusion can be studied for a fixed Zr disorder.

\begin{figure*}[t]
\centering
\includegraphics[width = \linewidth]{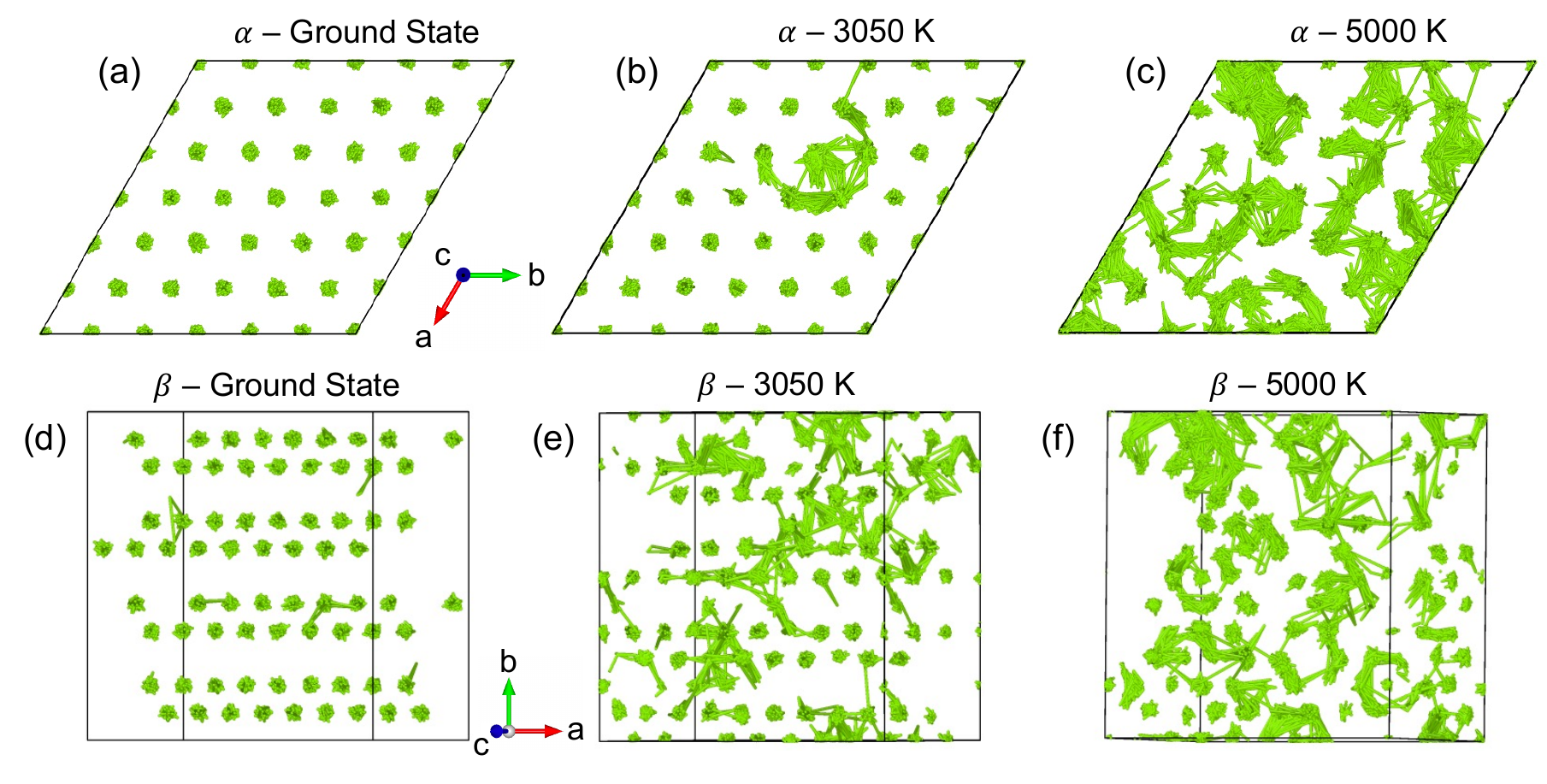}
\caption{The RT ($T=300$ K) Li displacements every 5 ps from the MD simulations of (a)-(c) $\alpha$-LZC and (d)-(f) $\beta$-LZC. The different degrees of Zr disorder are indicated by the effective temperatures with which they are obtained in MC.
}
\label{fig:traj_300K}
\end{figure*}

Figure \ref{fig:landscape_MD}c--d displays the Li diffusivities at temperatures from $T=300$ K to $700$ K in these structures. The ground-state structures (no Zr disorder) are represented by red squares, structures with partial Zr disorder ($T_{\text{eff}}=3050$ K) are represented by orange dots, and structures beyond the Zr order-disorder transition ($T_{\text{eff}}=5000$ K) are represented by green triangles. The dashed lines in Figure \ref{fig:landscape_MD}c--d are fits to the activation energies using the diffusivities obtained from the low-temperature range ($< 500$ K). The unfilled squares represent diffusivities for which a limited amount of Li hopping is observed in the MD simulations leading to high error bars on the diffusivity (see Figure \ref{fig:traj_300K}a and d as examples).

For structures beyond the Zr order--disorder transition ($T_{\text{eff}}=5000$ K, green triangles), the dashed line fit from the low-temperature region closely aligns with the high-temperature simulated diffusivities (500 K to 800 K), which indicates that the diffusion coefficient satisfies a simple Arrhenius law over the full temperature region. In contrast, structures without Zr disorder (ground state, red squares) exhibit marked non-Arrhenius behavior for both \(\alpha\)- and \(\beta\)-LZC, as they show no overlap between the dashed lines and high-temperature diffusivities. The steep slope of these red dashed lines indicates high activation energies near RT ($E_a = 0.97\pm 0.08$ eV for \(\alpha\)-LZC and $E_a = 0.68 \pm 0.04$ eV for \(\beta\)-LZC), significantly larger than those determined by nudged-elastic band (NEB) calculations within the hcp and ccp halide framework \cite{wang2019lithium}. The analysis derived from MD simulations aligns closely with the thermodynamic data. The non-Arrhenius behavior of Li transport in the structures with ground-state Zr ordering arises from the order--disorder transition of Li and vacancies near 500 K (see blue lines in Figure \ref{fig:landscape_MD}a--b), whereas such a transition is not observed for Zr-disordered structures.

The $\alpha$- and $\beta$-LZC polymorph show distinct behavior when Zr is partially disordered ($T_{\text{eff}}=3050$ K, orange dots in Figure \ref{fig:landscape_MD}c--d). The diffusivity of Li in \(\alpha\)-LZC with $T_\text{eff}=3050$  K has an activation energy $E_{a}=0.42 \pm 0.02$ eV, which is in between that of the ground state ordering and fully disordered Zr. In contrast, for $\beta$-LZC, Li diffusion for partial Zr disorder ($T_{\text{eff}}=3050$ K) occurs with a similar activation energy than in the highly disordered case ($E_{a} = 0.31 \pm 0.02$ eV), suggesting that the ccp structures are conducive to Li diffusion even when Zr is only partially disordered. We find that the difference of activation energy in $\alpha$ and $\beta$-LZC is also reflected in the energy landscape, with \(\alpha\)-LZC showing some flattening of the energy landscape (orange line in Figure \ref{fig:landscape_MD}a). This mirrors the higher activation energy ($E_a = 0.42$ eV) as compared to the fully Zr-disordered state ($E_a = 0.28$ eV), as shown in Figure \ref{fig:landscape_MD}c. In contrast, the energy landscape for \(\beta\)-LZC (orange line in Figure \ref{fig:landscape_MD}b) closely resembles that of fully-disordered states (dark-red lines in Figure \ref{fig:landscape_MD}b), consistent with the MD findings that suggest similar activation energies ($E_a = 0.31$ eV) for the two degrees of Zr disorder in \(\beta\)-LZC.

An illustration of the difference in the Li transport kinetics is demonstrated in Figure \ref{fig:traj_300K}, where the RT Li displacements over 5 ps from MD simulations are plotted. Very limited movement of Li ions is observed within the Zr ground state structures, as the Li-ions oscillate around their equilibrium sites (Figure \ref{fig:traj_300K}a and c). At $T_{\text{eff}}=3050$ K, limited Li hops to nearest sites are observed for \(\alpha\)-LZC due to the partial Zr disorder, whereas \(\beta\)-LZC shows more extensive Li diffusion. With substantial Zr disorder ($T_{\text{eff}}=5000$ K), the RT Li displacements demonstrate interconnected diffusion pathways across Li sites, which rationalize the high Li mobility promoted by Zr disorder.

In the realm of ASSBs, chloride-based lithium superionic conductors stand out because of their high ionic conductivity, interface stability, and oxidation resistance. Optimizing the Li migration barriers and increasing Li site connectivity are important roads for enhancing Li-ion conductivity in chloride-based solid-state electrolytes \cite{Liu2020_tailoring}. Prior research has indicated relatively low energy barriers ($<$ 0.3 eV) for lithium diffusion in both hcp and ccp frameworks of chlorides from either NEB calculations or by fitting the diffusivities derived from AIMD simulations at high temperatures \cite{wang2019lithium}. However, the LZC synthesized by solid-state heating exhibits a much higher activation energy and lower Li-ion conductivity at RT than would be expected from these basic models. Ball-milled LZC, on the other hand, is more conductive.

We selected the LZC system as a representative model to elucidate the effect of cation disorder on ionic transport because it exists in both $\alpha$- (hcp) and $\beta$- (fcc) phases, encompassing the typical anion arrangements common for closed-packed halide conductors. We quantified the degree of cation disorder within LZC by distinguishing between two key degrees of freedom: Li/vacancy disorder and Zr disorder. Our CE-MC simulations reveal three key findings: (1) no spontaneous Zr disorder is attainable through thermal fluctuations alone at any temperature reasonably accessible in solid-state synthesis, (2) thermal fluctuations at RT are inadequate to sustain Li/vacancy disorder when Zr is well-ordered in its ground state, and (3) Zr disorder frustrates the configurational energy landscape and enables facile Li/vacancy disorder at RT. These findings are supported by MLIP-MD, which reveal significant non-Arrhenius behavior of the Li diffusivity when Zr is well-ordered in contrast to the Zr-disordered states, which show typical Arrhenius behavior in the full temperature range. Using MLIP-MD, evidence for non-Arrhenius behavior has also been shown for several other chloride compounds, such as Li$_3$YCl$_6$ \cite{Wang2023_frustration, Qi2021_gap} and Na$_{3-x}$Y$_{1-x}$Zr$_x$Cl$_6$ \cite{Wu2021_NaZrCl, Sebti2022_NaZrCl}, leading to reduced conductivities near room temperature.

Our simulations reveal that the \(\beta\)-LZC polymorph exhibits lower activation energy ($E_a=0.68$ eV) than $\alpha$-LZC ($E_a=0.97$ eV) when Zr is fully ordered and that $\beta$-LZC achieves an activation energy for Li-ion diffusion ($E_{a} = 0.31$ eV) similar to that of the highly disordered state even with partial Zr disorder (Figure \ref{fig:landscape_MD}d). While cation-disordered $\beta$-LZC has not been synthesized yet, the lower activation energy of $\beta$-LZC suggests that tuning cation occupancy may have the potential to achieve fast Li transport, thereby avoiding synthesis via high-energy ball milling. Experimentally, only alternative approaches such as aliovalent substitution (e.g., Sc/In) and lithium stuffing in the fcc framework have been used to achieve conductivities over $2$ mS/cm in solid-state heated materials \cite{kwak2022li+_CEJ, li2023high_CEJ}. A MLIP-MD simulation of the doped Li$_{2.5}$In$_{0.5}$Zr$_{0.5}$Cl$_{6}$ finds high Li diffusivity near RT (Figure S3), which agrees with the notion that creating available hopping sites enables Li percolation and improves ionic conductivity \cite{Liu2020_tailoring, Xiao2021_Li_stuff, Yu2023_science}.

In summary, our study identifies metal disorder as a critical determinant for achieving fast Li transport in Li$_2$ZrCl$_6$. The high effective temperature required for Zr disorder limits equilibrium pathways for its synthesis and highlights the importance of manipulating cation disorder through non-equilibrium synthesis strategies. As many such non-equilibrium synthesis approaches are often more expensive and less scalable, in particular given the air sensitivity of these materials, novel strategies, including structural or chemical modifications may be required. The application of doping strategies or the exploration of high-entropy approaches may serve as viable pathways to introduce cation disorder, thereby promoting the configurational entropy to facilitate the Li/vacancy disorder \cite{Zeng2022_science}. Finally, given the importance of metal disorder for Li transport, it may be required to better understand if such disorder remains thermally stable over the typical lifetimes expected for battery materials.

\begin{acknowledgement}
This work was supported by the Assistant Secretary for Energy Efficiency and Renewable Energy, Vehicle Technologies Office, under the Advanced Battery Materials Research (BMR) Program, of the U.S. Department of Energy under Contract No. DE-AC02-05CH11231. The computational modeling in this work was supported by the computational resources provided by the Extreme Science and Engineering Discovery Environment (XSEDE), supported by National Science Foundation grant number ACI1053575; the National Energy Research Scientific Computing Center (NERSC); and the National Renewable Energy Laboratory (NREL) clusters under ahlssic allocation. The authors thank Xiaochen Yang for valuable discussions and Yu Xie for the help on the Flare package setups.
\end{acknowledgement}

\begin{suppinfo}

Methodological descriptions of cluster expansions and machine learning interatomic potentials; disordered LZC structures; analysis of the Li site energies; MD simulations of Li$_{2.5}$In$_{0.5}$Zr$_{0.5}$Cl$_{6}$
\end{suppinfo}

\bibliography{references.bib}


\end{document}